\documentclass{appolb}
\usepackage{graphicx}
\usepackage{hyperref}

\begin{document}
\title{Real effective potentials for phase transitions in models with extended scalar sectors %
\thanks{Presented at Matter to the Deepest 2023 international workshop}%
}
\author{Károly Seller
\address{Institute for Theoretical Physics, ELTE Eötvös Loránd University,
Pázmány Péter sétány 1/A, H-1117 Budapest, Hungary}
}
\maketitle
\begin{abstract}
The effective potential is a widely used phenomenological tool to investigate phase transitions occurring in the early Universe at finite temperature.
In the standard perturbative treatment the potential becomes complex in some region of the background field values due to the non-convex nature of the classical potential in models with spontaneous symmetry breaking.
The imaginary part renders the minimization of the potential impossible when at finite temperature the absolute minimum is in the complex region. 
In this talk we introduce a simple method to calculate an effective potential that is fully real based on the optimized perturbation theory scheme.
We apply the method for models that extend the Standard Model with an additional singlet scalar.
\end{abstract}
  
\section{Introduction}

The effective potential defined via Legendre transformation in quantum field theories is the effective action evaluated for homogeneous field configurations.
By definition the exact effective potential is a convex and real function of the background field values.
While the exact potential may be obtained via non-perturbative methods, in phenomenological studies we use the loop expansion of the defining path integral and truncate the obtained series at some finite order.

The loop-expanded effective potential can be split into a sum of the classical potential and loop corrections in which the quantum effects include vacuum ($T=0$) and finite temperature contributions.
In a general theory with multiple scalar fields with background field values $v_k$, the one-loop effective potential can be written as
\begin{equation}
    \label{eq:Veff_general}
    V^{[1]}_{\mathrm{eff}}\big(\{v_k\},T\big) = V_{\mathrm{cl}}\big(\{v_k\}\big) + \sum_{i\in\mathbf{P}} \left[V^{(1)}_i\big(\{v_k\}\big) + V^{(1)}_{\mathrm{T},\,i}\big(\{v_k\},T\big)\right]\,.
\end{equation}
Here $\{v_k\}$ is the set of background field values, $T$ is the temperature, and ${\bf P}$ denotes the set of particles coupled to the scalar fields.
The vacuum quantum corrections are in general UV and IR divergent and they need to be renormalized using standard renormalization techniques.
The IR divergence arises due to the presence of massless Goldstone modes, and at one loop order it is removed by considering the pole mass parametrization scheme.

In the SM, the effective potential is a function of a single vacuum expectation value $v$ of the Brout-Englert-Higgs (BEH) field.
At zero temperature the finite one-loop effective potential is given in Ref.~\cite{Seller:2023xkn}.
Concerning the thermal evolution of $v$ the interesting parts are the scalar contributions
\begin{equation}
    \label{eq:SMeffpot}
    V^{[1]}_{\rm SM,eff}(v) \supset \frac{1}{64\pi^2}\left[m_h^4(v)\ln\frac{m_h^2(v)}{M_h^2}
    + 3 m_G^4(v)\ln\frac{m_G^2(v)}{M_h^2}\right]\,.
\end{equation}
Here $M_h^2$ is the Higgs boson mass and the background dependent scalar masses are given by $m_h^2(v)=\mu^2+3\lambda v^2$ and $m_G^2(v)=\mu^2+\lambda v^2$, where the square mass parameter is $\mu^2=-M_h^2/2<0$.
Note that the regularization scale dependence was removed by a finite wave function renormalization factor that fixed the residue of the Higgs boson propagator at the pole mass to unity.

The one-loop SM effective potential $V^{[1]}_{\rm SM,eff}(v)$ is complex for field values $v<v_0$, where $v_0=\sqrt{-\mu^2/\lambda}$ is the minimum (physical point) of the potential.
The reason for the appearance of imaginary parts is due to the scalar contributions of Eq.~(\ref{eq:SMeffpot}) that are complex when $m_h^2(v)<0$ or $m_G^2(v)<0$.

At high temperature the SM relaxes to its symmetric phase where the vacuum expectation value of the BEH field vanishes and the effective potential becomes convex.
However, in the perturbative treatment the vacuum part of the effective potential close to the origin is complex as indicated above.
Adding thermal contributions does not remove the imaginary part (in fact it makes things worse) thus at finite temperature the potential cannot be minimised.
Consequently, the effective potential obtained by the standard perturbative treatment cannot be used to study phase transitions at finite temperature.

In this contribution, I introduce the optimized perturbation theory (OPT) scheme of Ref.~\cite{Chiku:1998kd} as a simple tool to obtain an effective potential that is real for all values of the background field.
I use it to obtain a real effective potential at zero temperature in the SM and in the SM extended by one singlet scalar field.
Finally, I show finite temperature contributions and calculate the critical temperatures in the superweak extension of the SM \cite{Trocsanyi:2018bkm}.
This contribution is based on Ref.~\cite{Seller:2023xkn}.

\section{Optimized perturbation theory}

The OPT scheme of Ref.~\cite{Chiku:1998kd} is an established calculational tool in quantum field theory that introduces a modified perturbative series.
The method preserves Ward identities and renormalizability even at higher orders in the loop expansion as it preserves the symmetries of the Lagrangian.
Additionally, the OPT method can be used to obtain an effective potential that remains real for all values of the background fields even at higher temperatures where minimization in the standard approach is hindered by the appearing imaginary parts at $v<v_0$.

\subsection{Parametrization of the SM}

In the OPT scheme we introduce a new squared mass parameter $m^2$ that we assume to be positive.
By reorganization of the Lagrangian terms we write the original Lagrangian potential in the form
\begin{equation}
        \mathcal{L}\supset\mathcal{V}[\phi]\equiv\mathcal{V}_\mathrm{OPT}[\phi]=m^2|\phi|^2+\lambda|\phi|^4+(\mu^2-m^2)|\phi|^2\,,
\end{equation}
where we treat the last term as a counterterm that only contributes at loop level.
We may interpret $m^2$ as the tree level value of the mass squared parameter.
With the shifted mass parameter the classical potential is convex and consequently the scalar masses are real.
Since the Lagrangian is unchanged it follows that the exact result is unchanged, in particular it is independent of $m^2$.

The one-loop effective potential of the SM at zero temperature is
\begin{equation}
    V_\mathrm{SM,OPT}^{[1]}(v;\mu^2,m^2)= \underbrace{V_\mathrm{cl}(v;m^2) + \lefteqn{\overbrace{\phantom{\frac{\mu^2-m^2}{2}v^2 + V^{(1)}(v;m^2)}}^{\displaystyle {\rm one-loop}}}\frac{\mu^2-m^2}{2}v^2}_{\displaystyle =V_{\rm cl}(v;\mu^2)} + V^{(1)}(v;m^2)\,.
\end{equation}
The effective potential at any finite order of perturbation theory will depend on three unknown parameters, $\mu^2,m^2,$ and $\lambda$. 
The dependence on $m^2$ only appears in the one-loop correction.

The parametrization of the OPT potential is done by using the standard parametrization conditions for the minimum of the potential and the mass of the Higgs boson,
\begin{equation}
    \label{eq:OPT_condition_1}
    \left.\frac{\partial V_\mathrm{SM,OPT}^{[1]}(v;\mu^2,m^2)}{\partial v}\right|_{v=v_0} = 0\,, \quad \left.\frac{\partial^2 V_\mathrm{SM,OPT}^{[1]}(v;\mu^2,m^2)}{\partial v^2}\right|_{v=v_0} = M_h^2\,.
\end{equation}
These two conditions fix two out of the three parameters.

The third condition is called the {\it principle of least sensitivity} that requires that the potential in the physical point (minimum) at fixed order is least dependent on the choice of $m^2$,
\begin{equation}
    \label{eq:OPT_condition_2}
    \left.\frac{\partial V_\mathrm{SM,OPT}^{(1)}(v;\mu^2,m^2)}{\partial m^2}\right|_{v=v_0} = 0\,.
\end{equation}
There is no guarantee that these three conditions have real solutions for the parameters such that $m^2>0$.
However, it so happens that for the SM the system is solvable and we find
\begin{equation}
    m^2=69~094.6{\rm GeV}^2,\quad\lambda=0.12~861,\quad\mu^2=-8~847.85{\rm GeV}^2\,.
\end{equation}
Since $m^2>0$ this potential is fully real for all values of the vacuum expectation value $v$.

\subsection{Parametrization of singlet scalar extended SM}

\begin{figure}[ht]
\centerline{
\includegraphics[width=12.5cm]{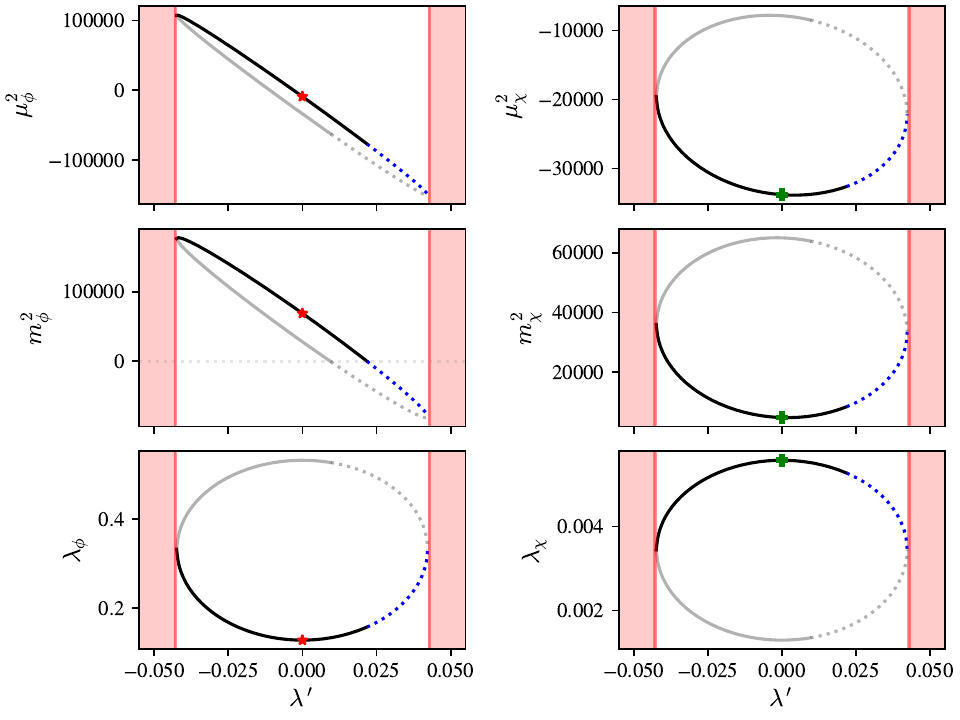}}
\caption{Parametrization of the SM extended with a singlet complex scalar field.
In the figure we used $M_s=260~$GeV and $w_0=10v_0$, with the regularization scale fixed at the top quark mass, $Q=M_t$.}
\label{Fig:F2H}
\end{figure}

In particle physics models that involve multiple scalar fields the effective potential is a function of multiple vacuum expectation values.
Thankfully, the generalization of the OPT scheme introduced above is trivial to models with any number of scalar fields.
The general prescription is to introduce shifted mass parameters $m_\alpha^2$ for all $\mu_\alpha^2<0$ present in the Lagrangian.

One of the simplest extensions is to add one singlet complex scalar field to the SM with vacuum expectation value $w\neq 0$.
Assuming a $Z_2$ symmetric potential and non-vanishing mixing between the scalar fields, one has:
\begin{equation}
    V_{\rm OPT}^{[1]}(v,w;\mu^2_{\phi},\mu^2_\chi,m^2_{\phi},m^2_\chi) = V_{\rm cl}(v,w;\mu^2_\phi,\mu^2_\chi) + V^{(1)}(v,w;m^2_\phi,m^2_\chi)\,.
\end{equation}
Similarly to the SM case, the classical potential and the counterterms combine to form the non-convex classical potential evaluated at $\mu_{\phi,\chi}^2<0$, while the one-loop contributions are all evaluated with $m_{\phi,\chi}^2>0$.

The parametrization conditions are easily generalized to the case of multiple scalar fields.
In the SM extended with a singlet scalar there are in total seven parameters: $\mu^2_\phi,\mu^2_\chi,m^2_\phi,m^2_\chi,\lambda_\phi,\lambda_\chi,$ and $\lambda'\,.$ 
Two conditions are used to fix the minimum of the potential:
\begin{equation}
    \label{eq:OPT1}
    \left.\frac{\partial V_\mathrm{OPT}^{[1]}(v,w)}{\partial v}\right|_{v=v_0,w=w_0} = \left.\frac{\partial V_\mathrm{OPT}^{[1]}(v,w)}{\partial w}\right|_{v=v_0,w=w_0} = 0\,.
\end{equation}
The condition to fix the scalar masses is modified, as the second derivative of the potential is now a symmetric matrix.
We set the eigenvalues of this curvature matrix to be equal to the scalar masses ($v_i=(v,w)_i$),
\begin{equation}
    \label{eq:OPT2}
    \left.\mathbf{M}^2_{ij}=\partial_{v_i}\partial_{v_j} V_\mathrm{OPT}^{[1]}\right|_{v=v_0,w=w_0} \longrightarrow \mathrm{diag}(M_h^2,M_s^2)\,.
\end{equation}

The last two conditions are the principle of minimum sensitivity enforced on both $m_\phi^2$ and $m_\chi^2$ parameters:
\begin{equation}
    \label{eq:OPT3}
    \left.\frac{\partial V_\mathrm{OPT}^{[1]}(v,w)}{\partial m_\phi^2}\right|_{v=v_0,w=w_0} = \left.\frac{\partial V_\mathrm{OPT}^{[1]}(v,w)}{\partial m_\chi^2}\right|_{v=v_0,w=w_0} = 0\,.
\end{equation}
The six conditions of Eqs.~(\ref{eq:OPT1})--(\ref{eq:OPT3}) fix all but one parameter of the one-loop effective potential.
We choose the scalar mixing parameter $\lambda'$ as the free paramater in terms of which we study the solution to the parametrization conditions for various values of $M_s^2$ and $w_0$.

An example parametrization is shown in Fig.~\ref{Fig:F2H} for a singlet scalar mass of $M_s=260~$GeV and vacuum expectation value $w_0=10v_0$.
The scalar mixing parameter $\lambda'$ is constrained by the mass difference of the scalar bosons to a symmetric interval around $\lambda'=0$, the excluded region is shown in red.
For any $\lambda'$ there are two solutions for each parameter: this merely reflects our ability to choose one or the other scalar boson as the Higgs boson.
The physical branch, containing the red and green points in Fig.~\ref{Fig:F2H} is chosen based on the requirement that the mass of the Higgs boson depends mostly on the vacuum expectation value of the BEH field.
In the figure, we also see that a solution with both $m_{\phi,\chi}^2>0$ is not always possible.
This is an artefact of the choice of regularization scale $Q$ and could be easily removed by increasing $Q$.
In practice, the larger the new scalar mass, the larger the regularization scale we should require in order to maintain a real potential.

\section{Phase transitions}

\begin{figure}[t]
\centerline{
\includegraphics[width=6.25cm]{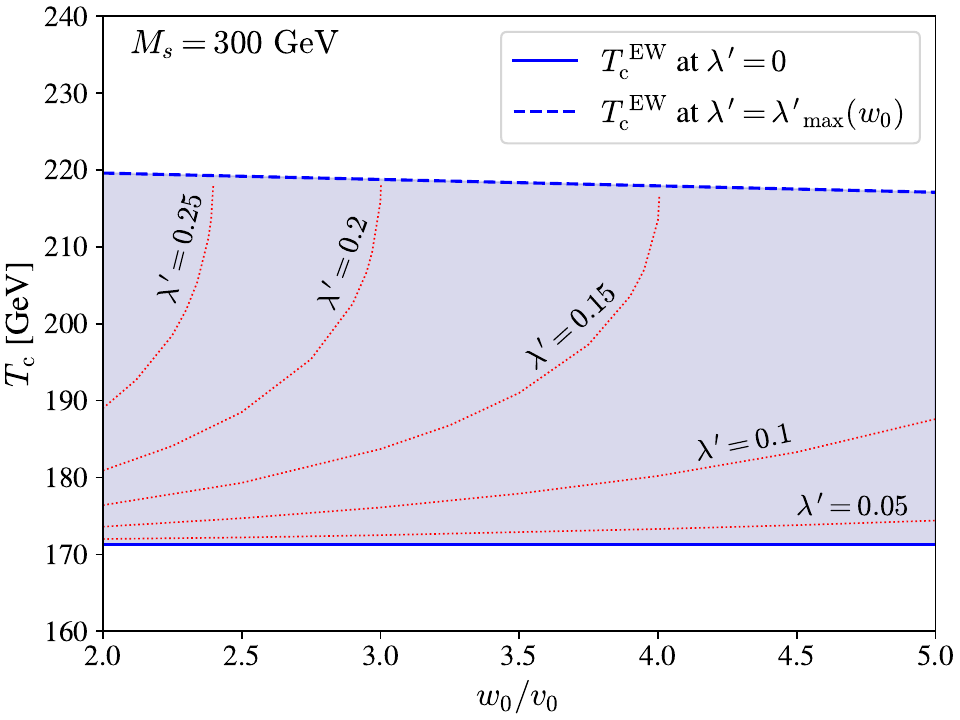}
\includegraphics[width=6.25cm]{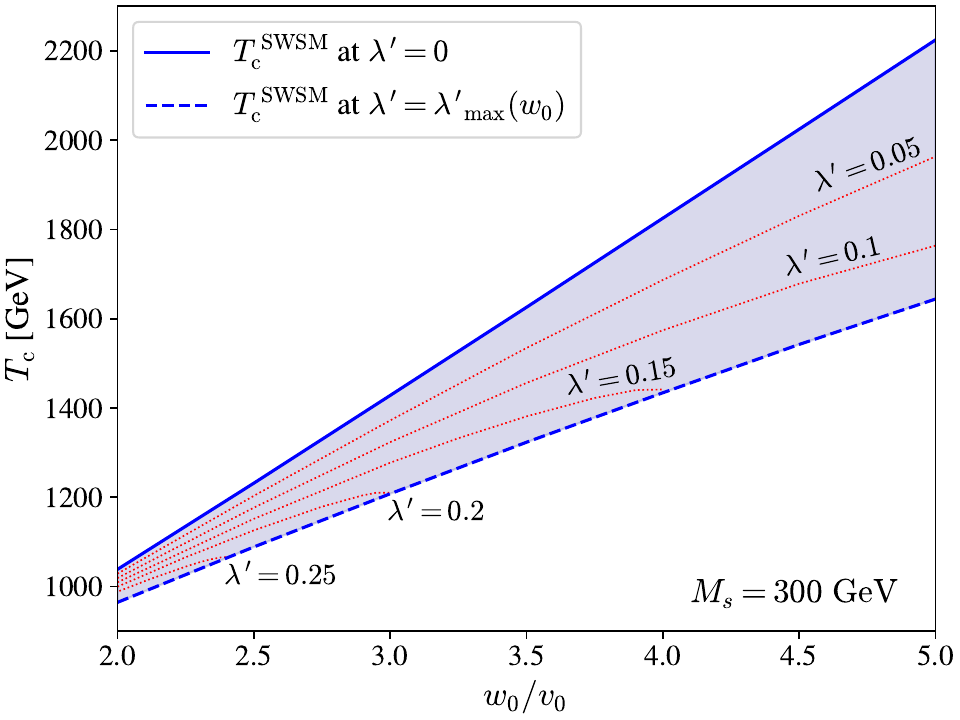}}
\caption{Critical temperatures of phase transitions in the superweak model.
In the left panel we see the temperatures where the SM BEH field acquires a non-zero vacuum expectation value.
In the right panel we see the temperatures where the new singlet complex scalar field acquires a non-zero vacuum expectation value.}
\label{fig:Tc}
\end{figure}

In this section, we apply the OPT scheme to calculate critical temperatures of phase transitions in the superweak model.
The finite temperature contributions to the potential do not introduce any new parameters, and we simply parametrize the potential at zero temperature as described in the previous section.

The thermal contribution of Eq.~(\ref{eq:Veff_general}) in the OPT scheme has the general form
\begin{equation}
    V_\mathrm{T,OPT}^{(1)}\Big(\{v_k\},\{m_\alpha^2\},T\Big)=\frac{T^4}{2\pi^2}\sum_{i\in\mathbf{P}} n_iJ^{(i)}_\pm\Big(m^2_i\big(\{v_k\},\{m_\alpha^2\}\big),T\Big)\,,
\end{equation}
where the fermionic or bosonic functions $J_\pm(m^2,T)$ are given in e.g.~Ref.~\cite{Arnold:1992rz}. 
Most importantly, they are real only if the particle masses are real.
This is achieved by the OPT parametrization and at first glance there is no new imaginary part introduced by the OPT thermal corrections.
We mention, however, that for $\lambda'<0$ the thermal mass of scalar fields can become imaginary at high temperatures due to $M^2_{\rm scalar}\supset \lambda' T^2<0$.
However, this result is related to the model definitions and not to the breakdown of perturbation theory, hence it is expected that the OPT scheme does not fix this.

In the superweak model the spectrum of the SM is extended with a new singlet scalar field and we may parametrize it as shown in the previous section.
In this model, the singlet scalar field acts as a Higgs boson in the dark sector, giving mass to a new gauge boson $Z'$ and to the right-handed neutrinos.
It follows that both the SM BEH field and the singlet scalar field necessarily acquire non-vanishing vacuum expectation values.
In general, the two scalar fields acquire vacuum expectation values at different temperatures that depend on the parameters of the model.
An example of the parameter dependence of the critical temperatures for $M_s=300~$GeV is shown in Fig.~\ref{fig:Tc}.

\section{Conclusions}

In this contribution, I have demonstrated how the optimized perturbation theory scheme can be used to calculate effective potentials in models with extended scalar sectors.
This method could be used to derive a potential that is real for all values of the background field and for all temperatures.
I applied the formalism to the SM extended by a singlet complex scalar, and calculated the critical temperatures of a two-step phase transition in the superweak model. 
The method could equally be used for more complicated models as well.

\begingroup\raggedright\endgroup

\end{document}